\documentclass[aps,prb, preprint, citeautoscript, showpacs,floatfix,superscriptaddress,showkeys]{revtex4-2}
\usepackage{bm}
\usepackage{amssymb}
\usepackage{graphicx}
\usepackage{amsmath}
\usepackage{textcomp}
\usepackage{colordvi}
\usepackage{multirow}
\usepackage{ulem}
\usepackage{tikz}
\usepackage{color}
\usepackage{makecell}
\usepackage{booktabs}
\usepackage{braket}
\usepackage{xcolor}
\usepackage{array}
\usepackage{float}
\usepackage{multirow}

\begin{document}

\title{Photogalvanic Effects in Surface States of Topological Insulators under Perpendicular Magnetic Fields} 

\author{Haoyu Li}
\affiliation{GPL Photonics Laboratory, State Key Laboratory of Luminescence Science and Technology, Changchun Institute of Optics, Fine Mechanics and Physics, Chinese Academy of Sciences, Changchun 130033, China.
}
\affiliation{University of Chinese Academy of Science, Beijing 100039, China.}

\author{Kainan Chang}
\email{knchang@ciomp.ac.cn}
\affiliation{GPL Photonics Laboratory, State Key Laboratory of Luminescence Science and Technology, Changchun Institute of Optics, Fine Mechanics and Physics, Chinese Academy of Sciences, Changchun 130033, China.
}

\author{Wang-Kong Tse}
\email{wktse@ua.edu}
\affiliation{Department of Physics and Astronomy, The University of Alabama, Alabama 35487, USA.}

\author{Jin Luo Cheng}
\email{jlcheng@ciomp.ac.cn}
\affiliation{GPL Photonics Laboratory, State Key Laboratory of Luminescence Science and Technology, Changchun Institute of Optics, Fine Mechanics and Physics, Chinese Academy of Sciences, Changchun 130033, China.
}
\affiliation{University of Chinese Academy of Science, Beijing 100039, China.}

\date{\today}

\begin{abstract}
We present a theoretical study of the nonlinear magneto-optical shift conductivity in the surface states of the prototypical topological insulator Bi$_2$Se$_3$ under a perpendicular quantizing magnetic field. 
By describing the electronic states as Landau levels and using a
perturbative approach, we derive the microscopic expression for the
shift conductivity $\sigma^{(2);\alpha\beta\gamma}(-\omega,\omega)$,
where  $\alpha,\beta,\gamma=\pm$ stand for the 
circular polarization of light
and $\omega$ is the light frequency; the spectra are
further decomposed into contributions from the interband and intraband optical transitions, for which the selection rules are identified.
Considering that the system possesses $C_3$ point group of symmetry,  the nonzero components of the conductivity tensor are $\sigma^{(2);-++}=[\sigma^{(2);+--}]^\ast$. Therefore, a pure circularly polarized light generates zero shift current.  In the clean
limit, the conductivities are nonzero only for 
discrete
photon energies  because of the discrete Landau levels and energy conservation,  and they become Lorentzian lineshapes with the inclusion of damping, which relaxes the condition of energy conservation. The dependence of the spectra on the damping parameters, the magnetic fields, and the chemical potentials 
is
investigated in detail. Our results reveal that the shift current is highly tunable by the chemical potential and the magnetic field. These results underscore the potential of topological insulators for tunable, strong nonlinear magneto-optical applications.

\end{abstract}

\keywords{shift current, topological insulator, surface states, Landau level, selection rules, nonlinear magneto-optical effect}

\pacs{42.65.-k, 73.20.At, 71.70.Di}

\maketitle 

\section{Introduction}

Three-dimensional topological insulators (TIs), known for their insulating bulk and conductive surface states, have attracted considerable interest due to their unique properties, such as spin-momentum locking and strong spin-orbit coupling \cite{PhysRevLett.114.257202, NatureMaterials13_699_2014, PhysRevLett.105.266806}. These features make TIs promising candidates for applications in optoelectronics, spintronics, and quantum computing.
In particular, the nonlinear optical response of TIs has been extensively studied both theoretically and experimentally on second-harmonic generation \cite{PhysRevB.86.035327, PhysRevB.91.195307, PhysRevB.99.155146}, third-harmonic generation \cite{ncomms11421}, high-harmonic generation \cite{NaturePhysics.17.311.2021, NanoLett.21.8970.2021},  and photogalvanic effects \cite{PhysRevLett.122.167401}.
Among various nonlinear optical phenomena, the photogalvanic effect -- which generates a  direct current under uniform illumination -- is of significant interest for applications in photodetection and solar energy conversion to surpass the Shockley-Queisser limit \cite{Zheng2023,Kaner2020,Cook2017}.
A prominent mechanism underlying this effect is the shift current, a second-order bulk photovoltaic effect originating from a real-space charge displacement of electron wavepackets during interband optical transitions \cite{JapaneseJournalofAppliedPhysics.63.060101.2024, PhysRevB.107.L201201}.
However, the bulk electronic states in TIs are centrosymmetric, and no shift current can be generated; instead, shift currents have been observed on TI surfaces \cite{NNANO.2011.214},  
where the inversion symmetry is broken \cite{PhysRevB.88.075144}.  
Theoretical studies have shown the tunablity of the shift current by applying strains  \cite{J.Phys.Chem.C.2024.128.13373_13378}, chemical potential \cite{PhysRevB.93.081403}, and the nanostructures of TI like nanowires \cite{Appl.Phys.Lett.116.172402.2020, Appl.Phys.Lett.117.262401.2020}. Experimentally,  asymmetric carrier distributions around the interface can lead to resonant shift currents, which is confirmed by the time- and angle-resolved photoemission spectroscopy \cite{PhysRevLett.122.167401}, highlighting the role of surface-state topology.

In addition, external magnetic fields provide an extra means to control photocurrent generation \cite{PhysRevB.107.L201201}.
For instance, in Bi$_2$Se$_3$, an in-plane magnetic field has been shown to induce a photogalvanic current through the interplay between orbital and Zeeman couplings \cite{PhysRevB.88.075144}. 
However, studies so far have primarily focused on in-plane field configurations, where the formation of Landau levels is suppressed. 
In contrast, our recent study shows that a strong perpendicular magnetic field -- which quantizes the electronic states into discrete Landau levels -- can have significant effects on second harmonic generation  \cite{PhysRevB.111.205408,PhysRevB.97.125417}; thus it is interesting to examine how the shift current can be affected by the quantizing Landau levels, which is the focus of this work.  Furthermore, considering that the shift current response requires real absorption of a photon to conserve the energy while second harmonic generation does not, it is also interesting to understand how the discrete Landau levels participate in the photogalvanic effects.

In this work, we address above gaps by presenting a comprehensive theoretical study on the shift
conductivity for the surface states of  Bi$_2$Se$_3$ under a perpendicular magnetic
field. 
From this, we derive an explicit expression for the second-order shift conductivity tensor $\sigma^{(2);\tau\alpha\beta}(-\omega,\omega)$ for circularly polarized light.
We show that the shift conductivity can be understood in terms
of optical transitions between Landau levels, governed by specific
selection rules. 
The dependence of the shift current on the chemical
potential and magnetic field strength is investigated,
providing insight into the tunability of nonlinear magneto-optical
responses in topological surface states.

This paper is organized as follows: In Sec.\,\ref{model} we present the theoretical model and the shift conductivity expression; in Sec.\,\ref{results} we analyze the selection rules, and then discuss results in different chemical potentials and magnetic fields; finally, we conclude in  Sec.\,\ref{conclusion}.
\section{Model}
\label{model}

\begin{figure}[!htp]
    \centering
    \includegraphics{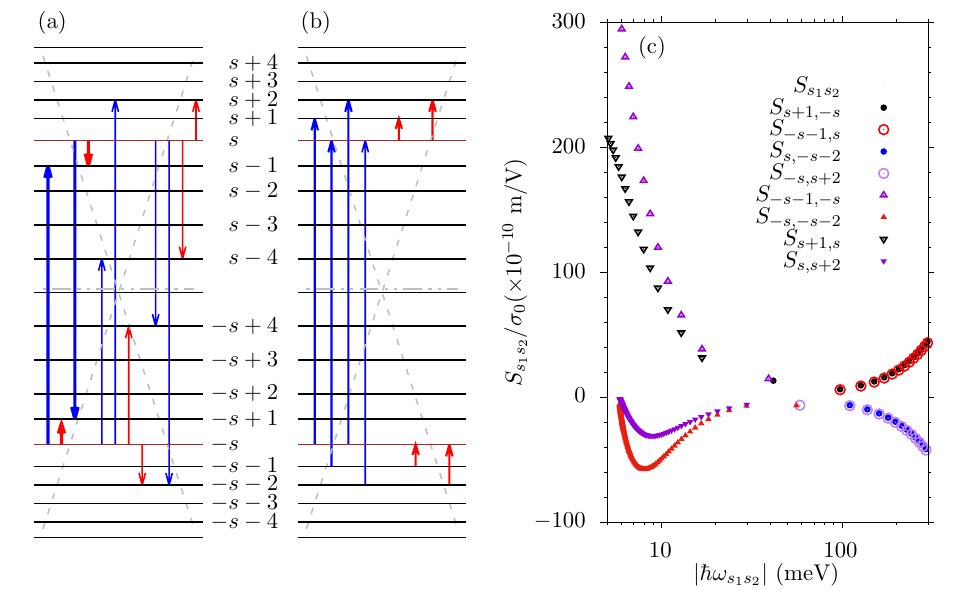}
    \caption{(a) Illustration of the selection rules for Berry connection $\xi^+_{s_1 s_2}$ with $|s_1|-|s_2|=3l+1$, where thick (thin) arrows correspond  $l=0$  ($l=\pm1$) with large (small) values. See details in Ref.\,\citenum{PhysRevB.111.205408}.
    (b) Illustration of the selection rules for coefficient $S_{s_1s_2}$ in shift conductivity  with $|s_1|-|s_2|=1, -2$, which is defined in Eq.\eqref{eq:S}. Note, both in (a) and (b), the blue and red arrows represent interband and intraband transitions, respectively.
      (c) All nonzero $S_{s_1 s_2}$ versus their respective discrete transition energies $|\hbar\omega_{s_1s_2}|$, where the Landau energies are calculated at $B=5$ T.
      $\sigma_0 = {e^2}/({4 \hbar})$.}
    \label{fig:band} 
\end{figure}

We study the optical response of the surface states for a topological
insulator Bi$_2$Se$_3$ with applying an external magnetic field $B$
along the $z$-direction. Without the magnetic field, the Hamiltonian  is  
$H(\bm p)=v_F(p_y\sigma_x-p_x\sigma_y)+\frac{\lambda}{2\hbar^3}(p_+^3 +
p_-^3)\sigma_z$, where 
$v_F$ is the Fermi velocity,
$\bm p$ is the momentum operator with $p_\pm =
p_x\pm i p_y$, 
$\sigma_{x,y,z}$ are Pauli matrices,
and the second term describes the hexagonal warping effects
with a strength parameter $\lambda$ \cite{Yar2022,PhysRevB.99.155146};
 with the magnetic field, the
Hamiltonian becomes $H(\bm p+|e|\bm A)+\Delta \sigma_z$, where 
$\bm A=(0,Bx,0)$ gives the vector
potential and $\Delta=g\mu_B
B/2$ determines the Zeeman coupling strength \cite{PhysRevB.111.205408}. 
The electronic states and Berry connections have been
well studied in our previous work \cite{PhysRevB.111.205408}, which
are briefly summarized as follows: 
First, the electronic states can be described by Landau levels $\Psi_s(\bm
r)$ with the Landau index $s=\cdots,-2,-1,0,1,2\cdots$, and the
corresponding eigenenergy is $\varepsilon_s$. Second, the Berry connection
between different Landau levels $s_1$ and $s_2$ is
$\bm\xi_{s_1s_2}$. For a circularly polarized incident light with a
polarization vector $\hat{\bm e}^\alpha=(\hat{\bm x}+i\alpha\hat{\bm
  y})/\sqrt{2}$ with $\alpha=\pm$, the light-matter  interaction is 
determined by $\xi_{s_1s_2}^\alpha=\bm\xi_{s_1s_2}\cdot\hat{\bm
  e}^\alpha$, where $\xi^\alpha_{s_1s_2}$ is a pure imaginary number and satisfies
$\xi_{s_1s_2}^+=\left[\xi_{s_2s_1}^-\right]^\ast$.
Without the warping term, and the eigenenergy can be given by $\varepsilon_s(0)=g_s \sqrt{\Delta^2+|s|(\hbar\omega_c)^2}$ with $g_s=\text{sgn}(s)$ for $s\neq0$ and $g_0=-1$, and the cyclotron energy $\hbar\omega_c=\sqrt{2\hbar eB}v_F$; the selection rule for $\xi^+_{s_1s_2}$ is
$|s_1|=|s_2|+1$. With the warping term, the correction to the eigenenergy is quite small when the Landau energies are less than
bulk gap of Bi$_2$Se$_3$ $\Delta_{\text{gap}}\approx300$ meV; and the selection rule for the Berry connections becomes
$|s_1|-|s_2|=3l+1$ for any integer $l$ and its value satisfies
$|\xi^+_{s_1s_2}|\propto \lambda^{|l|}$; up to the linear order of
the warping term, the values of $l$ are taken as $l=0,\pm
1$. 
While such selection rules could in principle be obtained from symmetry arguments, the complexity --  introduced by the presence of warping term and the choice of Landau gauge -- makes the wavefunctions difficult to handle within a purely symmetry-based analysis. Therefore, we adopt the more direct perturbation theory; see Ref. \citenum{PhysRevB.111.205408} for a detailed derivation.
Figure~\ref{fig:band} (a) illustrates the Landau levels and the allowed optical transitions for $\xi_{s_1s_2}^+$.
The underlying particle-hole symmetry is broken, which is crucial for the allowed second-order nonlinear optical response.

In this work, we focus on the direct charge currents generated by a light
pulse centered at a frequency $\omega$, for which the electric field is
$\bm E(t)=\bm E_0(t) e^{-i\omega t}+c.c.$  with the envelope function
$\bm E_0(t)$ slowly varying with time. 
Up to the second order of
the electric field, the direct current density $\bm J(t)$ can be written as
$\bm J(t)=\sum_{\alpha=\pm}J^\alpha(t)\hat{\bm e}^\alpha$ with  
\begin{align}
  J^\alpha(t)= 2\sum_{\beta,\gamma=\pm}
  \sigma^{(2);\alpha\beta\gamma}(-\omega,\omega)\left[E_0^{\overline{\beta}}(t)\right]^\ast
  E_0^\gamma(t)\,,
  \label{eq1}
\end{align}
where  $\overline{\alpha}=\mp$ for $\alpha=\pm$ 
($\overline{\beta}=\mp$ for $\beta=\pm$),
$J^\alpha(t)=\bm J(t)\cdot\hat{\bm e}^{\overline{\alpha}}$,  
$E_0^\alpha(t)=\bm E_0(t)\cdot\hat{\bm e}^{\overline{\alpha}}$, and $\sigma^{(2);\alpha\beta\gamma}(\omega,-\omega)$ the shift conductivity.
The microscopic expression of
$\sigma^{(2);\alpha\beta\gamma}(\omega,-\omega)$ is given
\cite{PhysRevB.111.205408} by
$\sigma^{(2);\alpha\beta\gamma}(\omega,-\omega)=\frac{1}{2}\left[\widetilde\sigma^{(2);\alpha\beta\gamma}(\omega,-\omega)
  +\widetilde\sigma^{(2);\alpha\gamma\beta}(-\omega,\omega)\right]$  with
\begin{align}
    \label{eq:sigma12A}
    \widetilde{\sigma}^{(2);\tau\alpha\beta}(-\omega,\omega)
    = \frac{ie^4B}{4\pi \hbar^2}\sum_{s_1 s_2 s}\frac{\hbar\omega_{s_1s_2}\xi^{\overline{\tau}}_{s_2 s_1} }{-\hbar\omega_{s_1 s_2} + i\Gamma_2}
    \left(\frac{\xi^{\alpha}_{s_1 s}\xi^{\beta}_{s s_2}f_{s_2 s}}{\hbar\omega - \hbar\omega_{s s_2} + i\Gamma_1}
    -\frac{\xi^{\beta}_{s_1 s}\xi^{\alpha}_{s s_2}f_{s s_1}}{\hbar\omega - \hbar\omega_{s_1 s} + i\Gamma_1}\right)\,.
\end{align}
Here $\hbar\omega_{s_1 s_2} =
\varepsilon_{s_1}-\varepsilon_{s_2}$, 
$f_{s_1 s_2} = f_{s_1} - f_{s_2}$ is the population difference where
$f_s = \theta(\mu-\varepsilon_s)$ is the
Fermi-Dirac distribution for a chemical potential $\mu$ at zero
temperature, and $\Gamma_{1,2}$ are the relaxation parameters.

The surface states of Bi$_2$Se$_3$ under a magnetic field $B$
possess 
symmetry of $C_3$ point group, which determines the nonzero
components of second order tensor
$\sigma^{(2);\tau\alpha\beta}(-\omega,\omega)$ as $\sigma^{(2);+--}(-\omega,\omega)$ and
$\sigma^{(2);-++}(-\omega,\omega)$ 
(see Appendix \ref{app} for symmetry analysis).  Furthermore, because $\bm J(t)$
is a real number, these two components also satisfy
$\sigma^{(2);-++}(-\omega,\omega)=[\sigma^{(2);+--}(\omega,-\omega)]^\ast=[\sigma^{(2);+--}(-\omega,\omega)]^\ast$,
then the dc current is
\begin{align}
  \bm J(t) &= 2\text{Re}\left[\sigma^{(2);-++}(-\omega,\omega)\hat{\bm
      e}^- \left(E_0^-(t)\right)^\ast E_0^+(t)\right] \notag\\
  &=
  \frac{|E_0(t)|^2}{\sqrt{2}}
  \left[\cos2\theta \begin{pmatrix}\sigma_r(\omega)\\\sigma_i(\omega)\end{pmatrix}+\cos\varphi
    \sin2\theta \begin{pmatrix} \sigma_i(\omega) \\ - \sigma_r(\omega)\end{pmatrix}\right]\,.\label{eq:jt}
\end{align}
Here we note $\bm E_0(t)=E_0(t)(\hat{\bm x}\cos\theta  +\hat{\bm y} \sin\theta
e^{i\varphi})$ with two parameters $\theta$ and $\varphi$ describing
its polarizaiton,
$\sigma_r(\omega)=\text{Re}\left[\sigma^{(2);-++}(-\omega,\omega)\right]$
and 
$\sigma_i(\omega)=\text{Im}\left[\sigma^{(2);-++}(-\omega,\omega)\right]$. It
is easy to find that a pure circularly polarized light ($\theta=\pi/4$
and $\varphi=\pm\pi/2$) cannot generate shift currents.


\section{Results}
\label{results}

\subsection{Selection rules}
To better understand the direct current response, we rewrite
$\sigma^{(2);-++}(-\omega,\omega)$ as
\begin{align}
\label{eq:sigma12AB}
\sigma^{(2);-++}(-\omega,\omega) &= \sum_{s_1 s_2}\left[
  {\cal A}_{s_1s_2} L_1(\hbar\omega-\hbar\omega_{s_1s_2})
+i{\cal B}_{s_1 s_2} L_2(\hbar\omega-\hbar\omega_{s_1s_2})
\right] f_{s_2s_1}\,,
\end{align}
where the coefficients are ${\cal A}_{s_1 s_2} = S_{s_2 s_1}-S_{s_1 s_2}$ and ${\cal B}_{s_1 s_2} = S_{s_1 s_2}+S_{s_2  s_1}$ with 
\begin{align}
    \label{eq:S}
    S_{s_1 s_2} =  \frac{ie^4B}{4\pi \hbar^2}\sum_{s}\xi^{+}_{s_2 s}
    \xi^{+}_{s s_1}\xi^{+}_{s_1 s_2}\left(\frac{1}{\hbar\omega_{s_2 s} + i\Gamma_2}-\frac{1}{\hbar\omega_{s s_1
        } + i\Gamma_2}\right)\,,
\end{align}
and two spectral functions
\begin{align}
  L_1(w)=   \frac{\Gamma_1^2}{w^2 +
    \Gamma_1^2}
  \,,\quad
  L_2(w) = \frac{w\Gamma_1}{w^2+\Gamma_1^2}\,.
\end{align}
From Eq.~\eqref{eq:sigma12AB}, the shift conductivity comprises the
contributions from transitions between Landau level pairs $(s_1,s_2)$. 
Each contribution is determined by two spectral functions,
$L_1(\hbar\omega-\hbar\omega_{s_1s_2})$ with 
amplitude ${\cal A}_{s_1s_2}$ and $L_2(\hbar\omega-\hbar\omega_{s_1s_2})$ with 
amplitude ${\cal B}_{s_1s_2}$, both centered at  $\hbar\omega=\hbar\omega_{s_1s_2}$; the Pauli blocking factor $f_{s_2s_1}$
is used to turn on/off the transition between the Landau level pair $(s_1,s_2)$
by adjusting the chemical potential, {\it i.e.}, the electron
density. In the following, we give a detailed analysis on Eq.~\eqref{eq:sigma12AB}:

\begin{enumerate}
\item We first discuss the selection rules of ${\cal A}$ and ${\cal B}$
by following Ref.~\citenum{PhysRevB.111.205408}. The
selection rule for $S_{s_1s_2}$ should satisfy 
$|s_1|-|s_2|=3l_1+1$, $|s_2|-|s|=3l_2+1$, and $|s|-|s_1|=3l_3+1$
simultaneously for integers $l_{1,2,3}$; thus $l_1+l_2+l_3=-1$. It 
can be found that the leading order of $S_{s_1s_2}$ is
$\propto\lambda$, for which $l_1$ can only take value of $0$ or $-1$,
then the selection rule for $S_{s_1s_2}$ becomes $|s_1|-|s_2|=1$ or
$-2$. 
This implies that contributions from $|s_1|-|s_2|=4$ transitions are of higher order in $\lambda$, making them negligible within the present treatment.
Further the selection rules for ${\cal A}_{s_1s_2}$ and ${\cal
  B}_{s_1s_2}$ are $|s_1|-|s_2|=\pm 1, \pm2$. Note that the selection
rules are different from the selection rules for second harmonic
generation (SHG) process of Landau levels. 
By taking the 
Landau levels $s\leq 0$
 with
$\varepsilon_s<0$ ($s>0$ with $\varepsilon_s>0$) as valence
(conduction) states, all transitions of ${\cal A}_{s_1s_2}$ and ${\cal
  B}_{s_1s_2}$ can be classified into the interband transitions, which occur between the conduction and valence bands, and
intraband transitions, which 
occur within
the conduction or valence
bands. All nonzero transitions with positive  
transition energies are listed in Table~\ref{tab:1} and shown in Fig.\ref{fig:band} (b), and the nonzero
values of $S_{s_1 s_2}$ are plotted in Fig.\ref{fig:band} (c). 

\item Because the values of the Berry connections $\xi_{s_1s_2}^+$ are pure imaginary,
thus $S_{s_1s_2}$ is approximately real for small relaxation parameter
$\Gamma_2$, further both ${\cal   A}_{s_1s_2}$ and  
${\cal B}_{s_1s_2}$ are approximately real; then the term involving
${\cal A}$ (${\cal B}$) in Eq.~\eqref{eq:sigma12AB} determines the
real (imaginary) part of
$\sigma^{(2);-++}$, which shows two types of spectral functions of 
$L_1(\hbar\omega-\hbar\omega_{s_1s_2})$ and $L_2(\hbar\omega-\hbar\omega_{s_1s_2})$ centered at photon energy
$\hbar\omega_{s_1s_2}$. In the clean limit as
$\Gamma_1\to0$, the function $L_1(\hbar\omega-\hbar\omega_{s_1s_2})$
is nonzero with values of $1$ only at
$\hbar\omega-\hbar\omega_{s_1s_2}$, which reflects the energy
conservation. 
However, this behavior is totally different from that in
the usual shift current in a crystal, where the energy
conservation lies in the Dirac-delta function
$\delta(\hbar\omega-\hbar\omega_{s_1s_2})$;  such a difference is 
induced by the discreteness of the Landau levels.  
It is also different from
the SHG in Landau levels, where a Dirac-delta
function \cite{PhysRevB.111.205408} also exists and the conductivity
 diverges as the photon energy approaches the
transition energy in the clean limit; such 
a difference indicates the
fundamentally distinct
physical processes 
underlying the shift current and SHG.
 In this case, $L_2=0$, thus the
shift current response exists only for discrete photon energies. For
finite $\Gamma_1$, $L_1$ shows a Lorentzian type lineshape
with a maximum $1$ located at $\hbar\omega=\hbar\omega_{s_1s_2}$;
while the spectral shape $L_2$ shows two extremes with values of
$\pm\frac{1}{2}$ located at
$\hbar\omega=\hbar\omega_{s_1s_2}\pm\Gamma_1$, and
$L_2(\hbar\omega-\hbar\omega_{s_1s_2})$ is always zero for
$\hbar\omega=\hbar\omega_{s_1s_2}$. Therefore, when the photon energy
matches the transition energy, $\sigma^{(2);-++}$ 
becomes purely real with
 $\sigma_i(\omega=\omega_{s_1s_2})=0$. Consequently, the shift current in
Eq.~(\ref{eq:jt}) 
simplifies to
\begin{align}
  \bm J(t) =
  \frac{|E_0(t)|^2}{\sqrt{2}}\sigma_r(\omega_{s_1s_2}) \begin{pmatrix}\cos2\theta
    \\ \cos\phi \sin2\theta\end{pmatrix}\,,
\end{align}
and its direction shows a definite relation with the light polarization.

\item For a small warping
parameter, $\varepsilon_s\approx-\varepsilon_{-s}$ for $s\ge1$; 
some
transitions have the same transition energies (they are
called as ``degenerate transitions''). 
For example, the
transitions of ${\cal
  A}_{s+2,-s}$ and ${\cal A}_{s,-s-2}$ are degenerate transitions. 
  However, they can be turned
on/off independently by adjusting the chemical potential $\mu$. By
defining $\mu_j$ as the chemical potential for which the highest occupied Landau level is $j$,
${\cal A}_{s+2,-s}$ is allowed only when the Fermi level is at $\mu_j$ with $j= s$ or $s+1$, whereas ${\cal A}_{s,-s-2}$ is allowed only at $\mu_j$ with $j= -s-2$ or $-s-1$.
 
\item 
The population difference
$f_{s_2s_1}$ for $s_1>s_2$ is not zero only for $\mu_j$ with $s_1\le j
<s_2$. A specific transition contributes to the conductivity only at
some filling factors: ${\cal A}_{1,0}$ only contributes
at $\mu_0$, while ${\cal A}_{2,0}$ can contribute at both $\mu_0$ and
$\mu_1$; similar cases can be found for all intraband
transitions. 
However, some interband transitions can contribute for many
chemical potentials. 
For example,  both ${\cal A}_{s+1,-s}$ and ${\cal A}_{s,-s-1}$
contribute for $\mu_j$ with $-s\le j\le s-1$, yielding a combined
transition amplitude (${\cal A}_{s+1,-s}+{\cal
  A}_{s,-s-1}$), while each additionally contributes at $\mu_s$ and $\mu_{-s-1}$, respectively.
 All these features are summarized in Table~\ref{tab:1}.

\end{enumerate}

\begin{table}[!ht]
\centering
\caption{All possible transitions for ${\cal A}_{s_1,s_2}$ and ${\cal
    B}_{s_1,s_2}$ for  $s_1>s_2$ with positive transition energies. 
   The column $X_{s_1s_2}$
stands for ${\cal A}$ or ${\cal B}$ with
$s\ge1$; the column ``value'' 
expresses
the values of $X$ as $S$ with the sign
$(-)$ for ${\cal B}$; the column ``$\hbar\omega_{s_1s_2}$'' is the
approximate transition energy for $X$, and the column ``$\mu_j$'' is the filling factor. ``DT'' means degenerate
transitions. }
{\small
\begin{tabular}{c|c|c|c|c|c|cc|c|c|c}
  \hline\hline		
  &\multicolumn{5}{c|}{interband transition}&&\multicolumn{4}{c}{intraband
    transition}\\
   \cline{2-11}
  &$X_{s_1s_2}$ & value & $\hbar\omega_{s_1s_2}$&\multicolumn{2}{c|}{$\mu_j$}&& $X_{s_1s_2}$ & value & $\hbar\omega_{s_1s_2}$&$\mu_j$\\
  \cline{2-11}
  &$X_{1,0}$ & $~~~~S_{1,0}$ & $\varepsilon_1-\varepsilon_0$ &/& $0$ && $X_{0,-2}$ &$~~~~S_{0,-2}$  & $\varepsilon_0+\varepsilon_2$&$-2,-1$\\
  \cline{2-11}
  &$X_{2,0}$ & $(-)S_{0,2}$ & $\varepsilon_2-\varepsilon_0$ &/& $0,1$
  && $X_{0,-1}$&$(-)S_{-1,0}$&$\varepsilon_0+\varepsilon_1$&$-1$\\
  \cline{1-11}
  \multirow{2}{*}{DT}&$X_{s+1,-s}$& $~~~~S_{s+1,-s}$ &
  \multirow{2}{*}{$\varepsilon_{s+1}+\varepsilon_s$} & \multirow{2}{*}{$[-s,s-1]$}&$s$ &    & $X_{s+1,s}$&$~~~~S_{s+1,s}$&\multirow{2}{*}{$\varepsilon_{s+1}-\varepsilon_s$}&$s$\\\cline{2-3}\cline{6-9}\cline{11-11}
  &$X_{s,-s-1}$& $(-) S_{-s-1,s}$ && &$-s-1$ &    & $ X_{-s,-s-1}$&$(-)S_{-s-1,-s}$&&$-s$\\\hline
  \multirow{4}{*}{DT}&\multirow{2}{*}{$X_{s+2,-s}$}& \multirow{2}{*}{$(-)S_{-s,s+2}$}
  &\multirow{4}{*}{$\varepsilon_{s+2}+\varepsilon_s$}&\multirow{4}{*}{$[-s,s-1]$}&$s$, &&\multirow{2}{*}{$X_{s+2,s}$} & \multirow{2}{*}{$(-)S_{s,s+2}$} & \multirow{4}{*}{$\varepsilon_{s+2}-\varepsilon_s$}& $s$, \\
  & & & & & $s+1$ && & & & $s+1$ \\
  \cline{2-3} \cline{6-9}\cline{11-11}
  & \multirow{2}{*}{$X_{s,-s-2}$} & \multirow{2}{*}{$S_{s,-s-2}$} & & & $-s-2$, && \multirow{2}{*}{$X_{-s,-s-2}$} & \multirow{2}{*}{$S_{-s,-s-2}$} & & $-s-2$, \\ 
  & & & & & $-s-1$ && & & & $-s-1$ \\
  \hline\hline		
\end{tabular}
}
\label{tab:1}
\end{table}

\subsection{Shift conductivity at $B = 5$ T}
Here we take $v_F = 5 \times 10^5$ m/s, $\lambda = 165 $
eV$\cdot$ \AA$^3$, $g = 8.4$, and $T = 0$ K
\cite{PhysRevB.111.205408}. 
Unless otherwise specified, the relaxation parameters are
$\Gamma_1=\Gamma_2= 1.3$ meV.

\begin{figure}[!htp]
    \centering
    \includegraphics[scale=1]{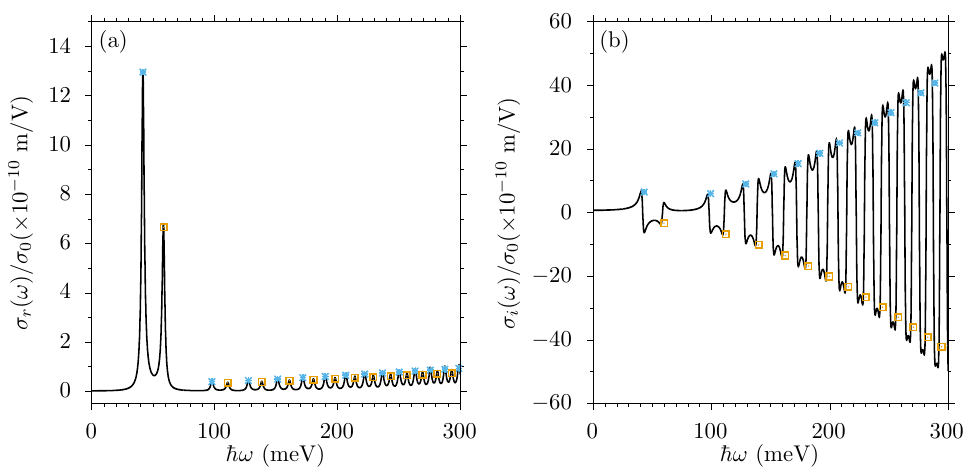}
    \caption{(a) 
The spectra of $\sigma_r(\omega)$ and the values of ${\cal A}_{s_1s_2}$ with $|s_1|-|s_2|=\pm1$ (star  symbols) and with $|s_1|-|s_2|=\pm2$ (square symbols). 
(b)
The spectra of $\sigma_i(\omega)$ and the values of ${\cal B}_{s_1s_2}$ with $|s_1|-|s_2|=\pm1$ (star  symbols) and  with $|s_1|-|s_2|=\pm2$ (square symbols). }
    \label{fig:5TS} 
\end{figure}

Figures~\ref{fig:5TS} (a, b) show the real and imaginary parts of the spectra
$\sigma^{(2);-++}$ under a magnetic field $B=5$~T and a chemical
potential $\mu=0$ eV (filling factor $\mu_0$).
According to Table~\ref{tab:1}, only interband transitions contribute to the
conductivity in this case, 
namely
$X_{1,0}$, $X_{2,0}$,
$X_{s+1,-s}+X_{s,-s-1}$, and $X_{s+2,-s}+X_{s,-s-2}$ for $s\ge1$. The spectra
of $\sigma_r(\omega)$ in Fig.~\ref{fig:5TS} (a) 
exhibit
discrete peaks, which are determined by ${\cal A}_{1,0}$, ${\cal A}_{2,0}$,
${\cal A}_{s+1,-s}+{\cal A}_{s,-s-1}$, and ${\cal A}_{s+2,-s}+{\cal
  A}_{s,-s-2}$ for $s\ge1$, 
  located at the transition
energies
$\hbar\omega_{1,0}\sim41.8$ meV, $\hbar\omega_{2,0}\sim58.5$ meV,
$\hbar\omega_{s+1,-s}$, and $\hbar\omega_{s+2,-s}$ for $s\leq1$, respectively.
It is
interesting that the first two peaks are 
an
order of magnitude higher
than subsequent ones.
This is because the former originate from
individual transitions, 
while the latter
are induced by degenerate
transitions (${\cal
  A}_{s+1,-s}$ and ${\cal A}_{s,-s-1}$, or ${\cal A}_{s+2,-s}$
and ${\cal A}_{s,-s-2}$ for $s\ge 1$), which have 
comparable amplitudes but opposite signs, leading to a significant cancellation.
The spectra of $\sigma_i(\omega)$ in Fig.~\ref{fig:5TS} (b) also show
discrete lineshapes formed by 
$L_2(\hbar\omega-\hbar\omega_{s_1s_2})$. 
Different from
$\sigma_r(\omega)$, the degenerate transitions ${\cal B}_{s+1,-s}+{\cal B}_{s,-s-1}$ and ${\cal B}_{s+2,-s}+{\cal
  B}_{s,-s-2}$ do not cancel each other. Therefore, with the photon
energy increases, the peak values increase as well. 

\subsection{Relaxation energy dependence}

\begin{figure}[!htp]
    \centering
    \includegraphics[scale=1]{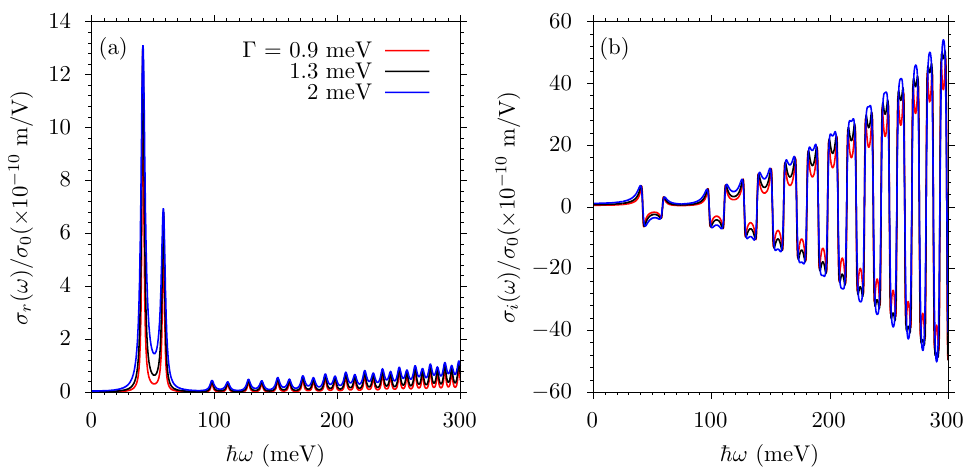}
    \caption{Spectra of  (a) $\sigma_{r}(\omega)$ and (b) $\sigma_{i}(\omega)$ for
     different relaxation parameters $\Gamma$.}
    \label{fig:Gamma}
\end{figure}

Figure \ref{fig:Gamma} shows the spectra of the conductivity
$\sigma_{r/i}(\omega)$ for  different relaxation 
parameters
$\Gamma_1=\Gamma_2=0.9$, $1.3$, and $2$ meV.  For $\sigma_r(\omega)$, it can be seen that with the
increasing $\Gamma$, both the peak location and the peak
value remain essentially unchanged, 
but the peak width increases significantly.  The peak
value hardly changes because the maximal value of $L_1(w)$ is
independent of the relaxation energy. Such relaxation parameter
insensitivity indicates that $\sigma_r(\omega)$ does not include the
contribution from the injection current process, and
$\sigma^{(2);-++}(\omega)$ is the shift conductivity. 
In contrast, for $\sigma_i(\omega)$, both the locations and values of these extremes
are slightly changed. 
The shifts of locations 
are because they depend on 
$\Gamma$, but the changes 
in the
extreme values are induced by $\Gamma$-dependent
${\cal A}_{s_1s_2}$. 
Notably, the insensitivity of the peak values of the shift current to $\Gamma$ contrasts with many conventional nonlinear magneto-optical effects  (e.g., injection current or SHG) \cite{PhysRevB.111.205408,Meng2023,Boyd2020}, where the response strength strongly depends on the relaxation parameter. This distinction shows the unique shift current in topological surface states under quantizing magnetic fields.

\subsection{Chemical potential dependence}

\begin{figure}[!htp]
    \centering
    \includegraphics{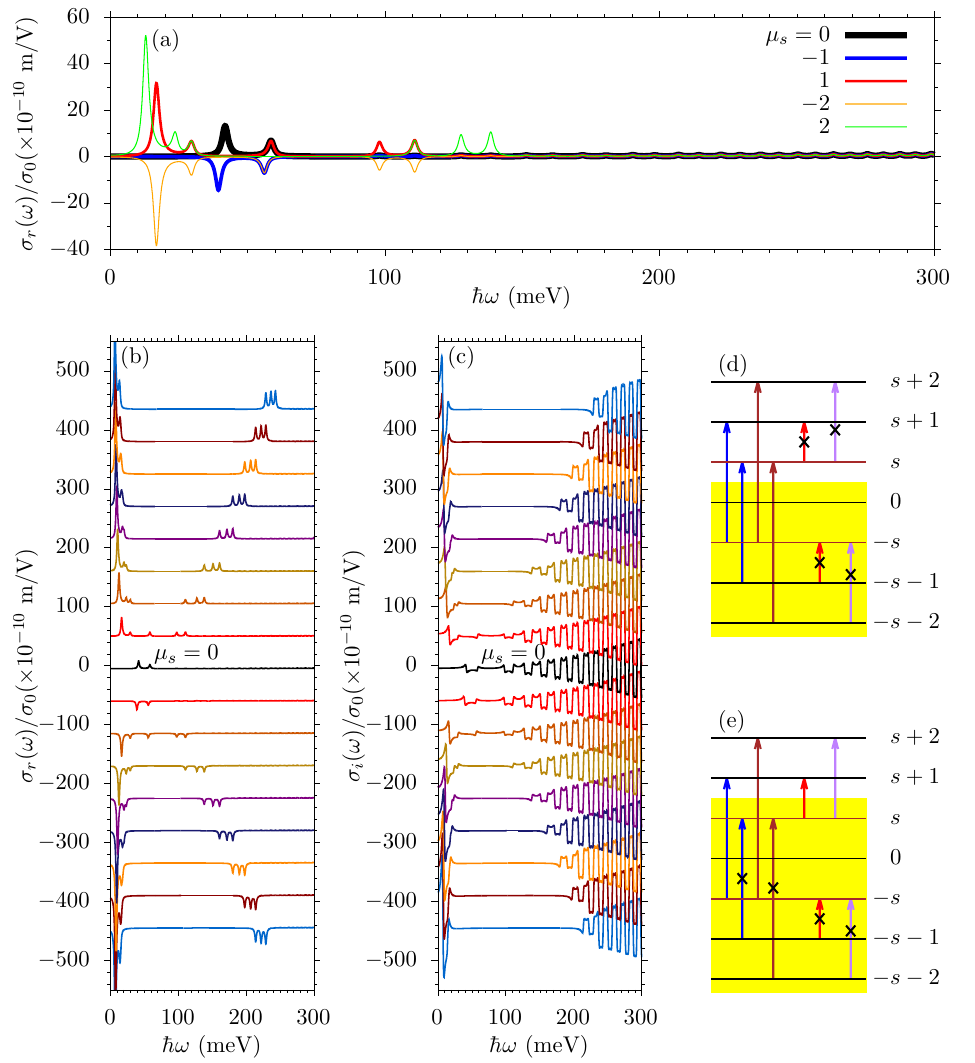}
    \caption{Shift conductivities with different $\mu_s$. (a) displays the spectra of $\sigma_{r}(\omega)$ for $\mu_s=0, \pm1, \pm2$. (b, c) show the spectra of  $\sigma_{r}(\omega)$ and $\sigma_{i}(\omega)$, respectively, where the black curve denotes $\mu_0$, and other curves above (below) it correspond to increasing  (decreasing) values of $\mu_s$. 
Illustration of the selection rules for the coefficient $S_{s_1s_2}$ when the highest occupied Landau level is (d) $0$ and (e) $s$ with $s > 0$, corresponding to the filling factors $\mu_0$ and $\mu_s$, respectively. Arrows of the same color denote a pair of degenerate transitions (see Table~\ref{tab:1}), while crosses mark forbidden transitions.}
    \label{fig:fxxxp} 
  \end{figure}

Now we study how the shift conductivity is affected by the chemical
potential $\mu$, or filling factor $\mu_s$. As an example,
Fig.~\ref{fig:fxxxp} (a) shows the spectra of $\sigma_{r}(\omega)$ for
different $\mu_s$ for $-2\le s\le 2$. Compared to the case for
$\mu_0$, the difference of
the spectra at different $\mu_s$ arises from the Pauli blocking
effects: (1) Some of the interband
transitions are blocked and the corresponding peaks disappear, and
lowest interband transition energy increases for $\mu_s$ with larger
$|s|$. Taking $\mu_2$ as an example, from Table~\ref{tab:1}, all available interband transitions are
${\cal A}_{s+1,-s}+{\cal A}_{s,-s-1}$ and ${\cal A}_{s+2,-s}+{\cal
  A}_{s,-s-2}$ for $s\ge3$, ${\cal A}_{3,-2}$, ${\cal A}_{3,-1}$, and
${\cal A}_{4,-2}$. The first interband transition starts from larger
energy $\hbar\omega_{3,-1}$. Compared to the case for $\mu_0$, there
disappear five interband transitions of ${\cal A}_{1,0}$, ${\cal A}_{2,0}$,
${\cal A}_{2,-1}$, ${\cal A}_{1,-3}$, and ${\cal A}_{2,-4}$, where the
latter three breaks the degenerate transition and induce relative large
peaks located at 
$\hbar\omega_{2,-1}$, $\hbar\omega_{1,-3}$, and $\hbar\omega_{2,-4}$.
(2) There appears intraband transitions. For $\mu_2$, they are
${\cal A}_{3,2}$, ${\cal A}_{4,2}$, and ${\cal A}_{3,1}$, which
correspond to lower transition energies, but with large
amplitudes. For $\mu_s$ with large $s$, the first intraband transition
energy is $\hbar\omega_{|s|+1,|s|}$, which decreases with increasing
$|s|$. The intraband transition shows analogous to the Drude
contribution. Similar analysis can be applied for other  
$\mu_s$. Figure~\ref{fig:fxxxp} (b, c) show the spectra of 
$\sigma_{r}(\omega)$ and $\sigma_i(\omega)$ for more $\mu_s$ with
$-8\le s\le 8$. For higher doped system, the predominant responses
occur in the lower photon energy regime induced by the intraband
transitions and in the higher photon energy regime induced by interband
transitons; no shift current
response exists between these two energy regimes.
Schematic diagrams illustrating the allowed and forbidden transitions for $S_{s_1,s_2}$ are presented in Figs.~\ref{fig:fxxxp} (d) and (e), where the filling factors are $\mu_0$ and $\mu_s$ ($s > 0$), respectively.
For $\mu_0$, both interband transitions within each degenerate pair are allowed, and no intraband transition occurs.
In contrast, for $\mu_s$, only one transition in each interband degenerate pair is allowed -- the other is Pauli-blocked, and among intraband transitions, only those near to the Fermi level become active.

\subsection{Magnetic field dependence}

\begin{figure}[!htp]
    \centering
    \includegraphics{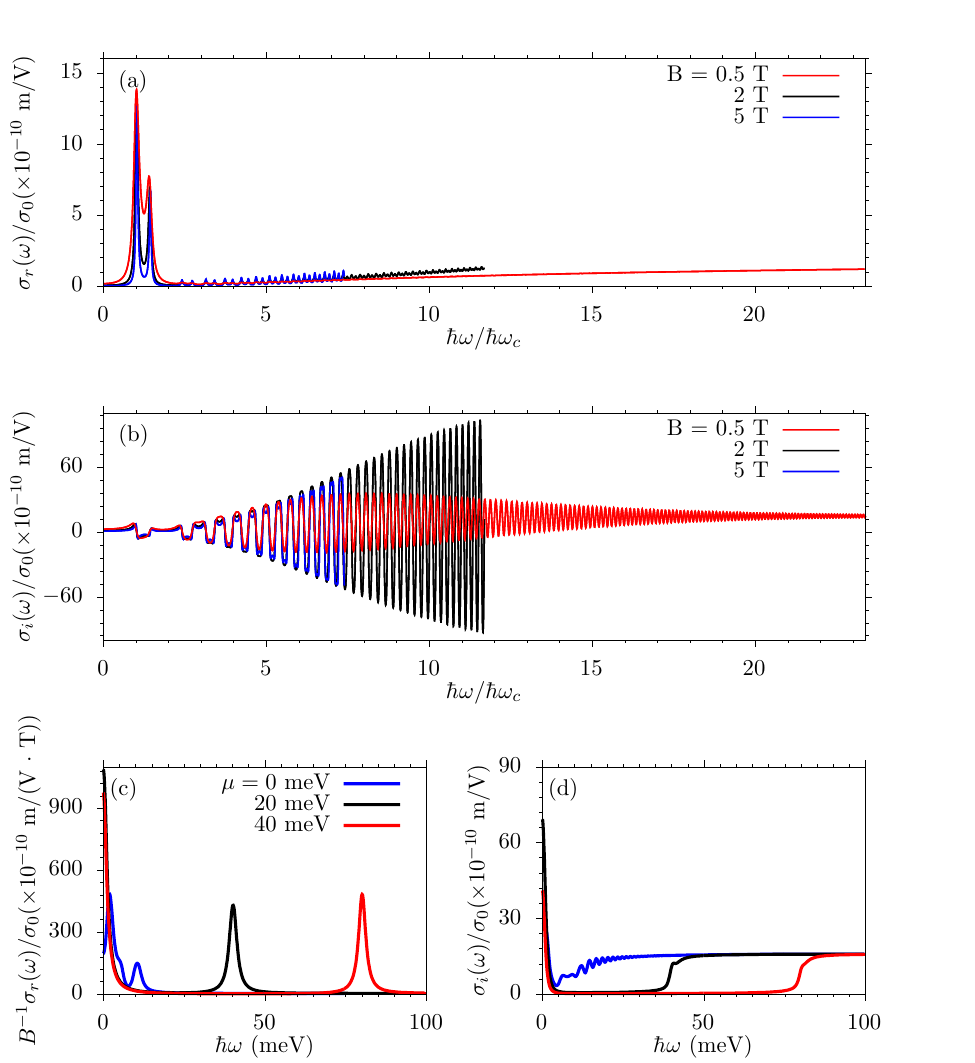}
    \caption{The spectra of (a) $\sigma_r(\omega)$ and (b)
      $\sigma_i(\omega)$ as functions of $\hbar\omega/\hbar\omega_c$
      for $\hbar\omega<300$~meV at $\mu=0$ eV and different magnetic
      field $B=0.5$, $2$, and $5$~T. 
     The spectra of (c) $B^{-1}\sigma_{r}(\omega)$  and (d) $\sigma_{i}(\omega)$ with different chemical potentials $\mu=0$, $20$,
      and $40$~meV for $B = 0.05$~T.}
    \label{fig:magnetic_field} 
  \end{figure}

Then we turn to the magnetic field dependence of 
$\sigma^{(2);-++}$. Figures~\ref{fig:magnetic_field} (a, b)
show the spectra of $\sigma_{r}$ and $\sigma_i$ as functions of
$\hbar\omega/\hbar\omega_c$ for $\hbar\omega<300$~meV at $\mu=0$ eV under $B=0.5$, $2$, and
$5$~T. From the previous analysis, as the magnetic field increases,
the cyclotron energy $\hbar\omega_c$ increases, and the energy
differences between Landau levels also increase; therefore, the peak
positions of $\sigma_{r/i}$ shift to higher energies and the intervals
between peaks also increase. With the energy scaling factor
$\hbar\omega_c$, it is obvious that all peaks and
oscillations are located approximately at the same values of
$\hbar\omega/\hbar\omega_c$.  
At $B = 0.5$ T, 
the peak intervals of $\sigma_r$ at
$\hbar\omega/\hbar\omega_c>8.3$ are smaller than the scaled damping parameters
$\Gamma/\hbar\omega_c\sim 0.10$, and then the neighboring peaks 
merge
into a smooth curve; however, $\sigma_i$ still show
oscillations, 
although their amplitude is greatly reduced and gradually vanishes at higher photon energies.
In fact, with further
decreasing the magnetic field to $B = 0.05$ T, $\sigma_i$ becomes a smooth curve as
well at larger photon energy, as shown in
Fig.~\ref{fig:magnetic_field} (c).

To better understand the behavior of the conductivity as the magnetic
field goes to zero, we can refer to  a symmetry analysis in the limit of  $B=0$ T, where $\sigma_{r}(\omega)$ 
vanishes \cite{PhysRevB.111.205408} but
$\sigma_i(\omega)$ should exist. For small magnetic
field, our numerical results show that $\sigma_{r}(\omega)$ is
proportional to $B$, and $B^{-1}\sigma_r(\omega)$ for $B=0.05$~T are shown in
Fig. \ref{fig:magnetic_field} (c) for $\mu=0$, $20$, and $40$~meV. The
spectra show two regions: at the lower photon energies
$B^{-1}\sigma_r(\omega)$ decreases with the photon energy, indicating
the intraband contribution to the shift conductivities; there appears
a peak at $\hbar\omega\approx 2|\mu|$, indicating the contributions
from the interband transitions. At a small magnetic field,
$\sigma_i(\omega)$ is nonzero, and its spectra are also shown in
Fig.~\ref{fig:magnetic_field} (c) for chemical potential $\mu = 0,
20, 40$ meV.  The resonant peaks are associated with the chemical potential $\mu$, which is similar with second harmonic generation \cite{PhysRevB.111.205408}.   At the lower photon energy regime, $\sigma_i(\omega)$
show similar features as that of $B^{-1}\sigma_r(\omega)$, however,
the interband contributions approximately show a step function, where the values   at
$\hbar\omega>2|\mu|$ are approximately independent of
the photon energy.

\section{Conclusion}
\label{conclusion}
In summary, we have investigated the nonlinear magneto-optical
conductivity associated with the shift current in the surface states
of the topological insulator  Bi$_2$Se$_3$  under 
a perpendicular magnetic field.
By describing the electronic states as Landau levels and employing a perturbative approach, we derived the microscopic expression for the shift conductivity tensor.
It shows that the shift current can be generated 
when the light is not pure circularly polarized,
 and exihibit both $x$ and $y$ components, which arise from resonant optical transitions between
occupied Landau levels $s_1$ and unoccupied Landau levels $s_2$,
governed by the selection rules $|s_1|-|s_2|=\pm 1, \pm 2$.

We provide a detailed analysis of the contributions from intraband and
interband transitions to the conductivity spectra, along with their
dependence on damping parameters, chemical potentials, and magnetic
fields. 
Intraband transitions occur at lower transition
energies, whereas interband transitions contribute at higher energies. 
The damping parameters 
primarily affect the spectral linewidths without significantly altering peak amplitudes.
The chemical potential
significantly 
tunes
the contributions from intraband and interband
transitions: 
at larger chemical potentials, the spectral features 
from
interband transitions shift to higher photon energies, while those
from intraband transitions shift to lower photon energies, thereby
leaving a
photon energy region with 
negligible photocurrent response.

The peak positions scale approximately with cyclotron energy
$\hbar\omega_c\propto \sqrt{B}$, while the peak
amplitudes exhibit a more complicated dependence due to the interplay
between
the magnetic fields and the damping parameters. 
For large damping parameters or very small magnetic fields, the discrete spectral peaks merge into a smooth curve. 
Under very small magnetic fields, the $y$-component of
the current is approximately proportional to the magnetic field,
whereas the $x$-component forms a smooth curve with a resonant peak
occurring at a photon energy roughly twice the chemical potential.

When we take the thickness of the surface states as $d = \hbar v_F / \Delta_{\text{gap}} \approx9.4$ {\AA}, which is estimated as the depth of the surface wave functions penetrating into the bulk \cite{PhysRevB.86.035327}, 
the amplitude of 
effective bulk
shift conductivity $\sigma(\omega)$ can reach 326 $\mu$A/V$^2$ with $\hbar\omega \approx 297$ meV, $B = 5$ T and $\mu_s = 0$.
This value is of the same order of magnitude as that reported in GeSe (approximately $200$ $\mu$A/V$^2$) \cite{Rangel2017}.
It is noteworthy that our predicted conductivity can be significantly enhanced through higher doping levels or the stronger magnetic fields.
These findings indicate a relaxation dependent insensitive and highly tunable nonlinear magneto-optical response in topological insulators,
and suggest their potential as functional nonlinear materials for developing advanced magneto-optical devices.

\begin{acknowledgments}
Work in China was supported by National Natural Science 
Foundation of China Grants No. 12034003 (K.C. and J.L.C.). Work in U.S. was supported by the U.S. Department of Energy, Office of Science, Basic Energy Sciences under Early Career Award No. DE-SC0019326 (W-K.T.).
\end{acknowledgments}

\appendix
\section{Symmetry analysis for circularly polarized conductivity}
\label{app}

The effective Hamiltonian for the (111) surface of $\mathrm{Bi_2Se_3}$ including hexagonal warping respects the $C_{3v}$ point group, which contains threefold rotations $C_3$ about the trigonal axis and vertical mirror planes $M$ \cite{Fu2009,Yar2022,Hsieh2011}.
When a uniform magnetic field is applied along the $z$-axis, time-reversal symmetry is broken, and the vertical mirror symmetry with respect to the $yz$ plane is also broken. 
Consequently, the point-group symmetry of the surface is reduced to the pure rotational subgroup $C_3$, generated by a $2\pi/3$ rotation 
\begin{align}
R_{2\pi/3}:\; (x,y) \rightarrow \bigl(x\cos\tfrac{2\pi}{3}-y\sin\tfrac{2\pi}{3},\;x\sin\tfrac{2\pi}{3}+y\cos\tfrac{2\pi}{3}\bigr)\,.
\end{align}

We adopt the circular polarization vectors
\begin{align}
\hat{\bm{e}}^+ = \frac{\hat{\bm x}+i\hat{\bm y}}{\sqrt{2}},\qquad 
\hat{\bm{e}}^- = \frac{\hat{\bm x}-i\hat{\bm y}}{\sqrt{2}}\,,
\end{align}
which transform under $R_{2\pi/3}$, as
\begin{align}
R_{2\pi/3}\,\hat{\bm{e}}^\pm = e^{\pm i2\pi/3}\hat{\bm{e}}^\pm\,.
\end{align}
The dc photocurrent density generated at second order can be written as Eq.\eqref{eq1}.
The tensor component $\sigma^{(2);\alpha\beta\gamma}$ must be invariant under the symmetry operations of the system.
Under $C_3$ rotation, the product of field amplitudes transforms as
\begin{align}
\bigl[E^{\overline{\beta}}\bigr]^* E^\gamma \;\rightarrow\; 
e^{i(\beta+\gamma)\frac{2\pi}{3}} \,\bigl[E^{\overline{\beta}}\bigr]^* E^\gamma\,.
\end{align}
For the current $J^\alpha$ to remain invariant, the total phase factor of the term 
$\sigma^{(2);\alpha\beta\gamma}[E^{\overline{\beta}}]^*E^\gamma$ must be unity.
It implies
\begin{align}
e^{-i\alpha\frac{2\pi}{3}}=
e^{i(\beta+\gamma)\frac{2\pi}{3}}\,, 
\end{align}
or equivalently
\begin{align}
\alpha \equiv \beta+\gamma \pmod{3}\,.
\end{align}
Becase $\alpha,\beta,\gamma=+,-$ (represented numerically as $+1$ and $-1$), the only independent non-vanishing components are
\begin{align}
\sigma^{(2);+--}(-\omega,\omega)\,,
\quad
\text{and} 
\quad
\sigma^{(2);-++}(-\omega,\omega)\,,
\end{align}
with the relation
$\sigma^{(2);-++}(-\omega,\omega) = \bigl[\sigma^{(2);+--}(-\omega,\omega)\bigr]^*$.
All other components are identically zero under the $C_3$ point group.

\bibliographystyle{iopart-num.bst}
\bibliography{refs}

\end{document}